\documentclass[12pt]{article}
\usepackage{epsf}
\usepackage{amsmath,amssymb}

\usepackage[dvips]{graphicx}

\usepackage{comment}

\begin{document}

\newcommand{\lsim}{\stackrel{<}{_\sim}}
\newcommand{\gsim}{\stackrel{>}{_\sim}}

\newcommand{\rem}[1]{{$\spadesuit$\bf #1$\spadesuit$}}

\renewcommand{\thefootnote}{\fnsymbol{footnote}}
\setcounter{footnote}{0}

\begin{titlepage}

\def\thefootnote{\fnsymbol{footnote}}

\begin{center}

\hfill UT-13-23\\
\hfill IPMU13-0108\\
\hfill May, 2013\\

\vskip .75in

{\large \bf 

  Lepton Flavor Violations
  in High-Scale SUSY \\
  with Right-Handed Neutrinos

}

\vskip .75in

{\large 
Takeo Moroi$^{(a,b)}$,
Minoru Nagai$^{(a)}$
and Tsutomu T. Yanagida$^{(b)}$
}

\vskip 0.25in

\vskip 0.25in

$^{(a)}$
{\em Department of Physics, University of Tokyo,
Tokyo 113-0033, Japan}

\vskip 0.1in

$^{(b)}$
{\em Kavli IPMU (WPI), University of Tokyo, Kashiwa, Chiba 277-8568, Japan}

\end{center}
\vskip .5in

\begin{abstract}

  Motivated by the recent discovery of the Higgs boson at $m_h\simeq
  126\ {\rm GeV}$ and also by the non-observation of superparticles at
  the LHC, high-scale SUSY, where the superparticles are as heavy as
  $O(10)\ {\rm TeV}$, has been recently proposed.  We study lepton-flavor
  violations (LFVs) in the high-scale SUSY with right-handed neutrinos.
  Even if the slepton masses are of $O(10)\ {\rm TeV}$, the
  renormalization group (RG) effects on the slepton mass-squared matrix
  may induce large enough LFVs which are within the reach of future
  LFV experiments.  We also discuss the implication of the
  right-handed neutrinos on the electroweak symmetry breaking in such
  a model, and show that the parameter region with the successful
  electroweak symmetry breaking is enlarged by the RG effects due to
  the right-handed neutrinos.

\end{abstract}

\end{titlepage}

\renewcommand{\thepage}{\arabic{page}}
\setcounter{page}{1}
\renewcommand{\thefootnote}{\#\arabic{footnote}}
\setcounter{footnote}{0}

The recent results of the LHC experiments have provided important
information about the physics at the TeV scale.  Given the fact that
supersymmetry is a prominent candidate of the physics beyond the
standard model, the observed Higgs boson mass about $126\ {\rm GeV}$
\cite{Aad:2012gk, Chatrchyan:2012gu} together with the lack of any
observation of superparticles \cite{Aad:2012fqa, Chatrchyan:2012jx}
strongly suggests that the SUSY-breaking scale is likely higher than
$O(10)\ {\rm TeV}$ \cite{Okada:1990gg, Okada:1990vk, Ellis:1990nz,
  Haber:1990aw} in the minimal SUSY standard model (MSSM).  If this is
the case, it may be very difficult to discover directly superparticles
at the LHC.

The purpose of this letter is to show that, with the seesaw mechanism
 \cite{Yanagida:1979as,  GellMann:1976pg, Minkowski:1977sc}, the rates of
lepton-flavor-violation (LFV) processes may be within the testable
ranges even if the mass scale of the superparticles is very high as
suggested by the observed Higgs mass.  Indeed, for the case where the
off-diagonal elements of the slepton mass-squared matrices are sizable
compared to the diagonal elements, two of the authors have pointed out
that the rates of the LFV processes may be within the reach of future
experiments if the masses of superparticles are $O(10)\ {\rm
  TeV}$ \cite{Moroi:2013sfa}.  Here, we show that such a size of the
off-diagonal elements can be naturally generated by the
renormalization group (RG) effects if there exist heavy right-handed
neutrinos \cite{Borzumati:1986qx, Hisano:1995nq, Hisano:1995cp}.  For
a demonstration of our point, we assume the universal SUSY-breaking
soft masses for all chiral multiplets at the GUT scale.  Such an
assumption is also advantageous to avoid serious SUSY FCNC problems
\cite{Gabbiani:1996hi}; in particular, the SUSY contribution to the
$\epsilon_K$ parameter can be sufficiently suppressed in this
assumption.  We will see that, in the parameter region of our interest
(i.e., the region with the sfermion masses of $\sim 10\ {\rm TeV}$,
where the lightest Higgs mass becomes about 126 GeV),
$Br(\mu\rightarrow e\gamma)$ can be as large as $\sim
10^{-13}-10^{-14}$, which may be tested in future experiments.

We also show that the presence of the heavy right-handed neutrinos
enlarge the parameter space for the successful electroweak symmetry
breaking.  It is often the case that the electroweak symmetry breaking
becomes unsuccessful if the scalar masses are much larger than the
gaugino masses.  In particular, this is the case in the pure gravity
mediation model \cite{Ibe:2006de, Ibe:2011aa, ArkaniHamed:2012gw} with the universal
SUSY-breaking soft masses at the GUT scale; in such a model, only a
very narrow region of the gravitino mass, $m_{3/2}\simeq (300-1500)$
TeV, can induce the correct electroweak symmetry breaking if the
right-handed neutrinos do not exist \cite{Evans:2013lpa}. Here, we
find that the electroweak symmetry breaking becomes possible with
smaller value of $m_{3/2}$.

Let us first introduce the model of our interest.  As we have
mentioned, we consider SUSY model with right-handed (s)neutrinos.
Denoting the up-type Higgs, lepton doublet, and right-handed neutrino
as $H_u$, $l$, and $N^c$, respectively, the relevant part of the
superpotential for our analysis is
\begin{align}
  W = W_{\rm MSSM} + y_{\nu,Ij} N^c_I l_j H_u 
  + \frac{1}{2} M_{N,IJ} N^c_I N^c_J,
  \label{superpot}
\end{align}
where $W_{\rm MSSM}$ is the superpotential of the MSSM, and the
indices $I$ and $j$ are flavor indices which run $1-3$.  With the
above superpotential, the active neutrinos become massive in the
seesaw mechanism \cite{Yanagida:1979as, GellMann:1976pg,
  Minkowski:1977sc}; the mass matrix of the active neutrinos is given
by
\begin{align}
  {\cal M}_{\nu,ij} = 
  \left[
    U_{\rm MNS}^T {\rm diag} (m_{\nu_L,1},m_{\nu_L,2},m_{\nu_L,3}) 
    U_{\rm MNS} 
  \right]_{ij}
  = 
  \frac{1}{2}
  y_{\nu,Ii} y_{\nu,Jj} v^2 \sin^2\beta (M_{N}^{-1})_{IJ} ,
  \label{seesaw}
\end{align}
where $v\simeq 246\ {\rm GeV}$ is the vacuum expectation value (VEV)
of the Higgs boson, and $\beta$ is the angle parametrizing the Higgs
VEVs (with $\tan\beta$ being the ratio of the VEVs of up- and
down-type Higgs bosons).  Here, $m_{\nu_L,i}$ are mass eigenvalues of
active neutrinos.  From neutrino-oscillation experiments, we have
information about the mass-squared differences.  In our analysis,
concentrating on the case of hierarchical mass matrix for active
neutrinos, we adopt the following as canonical values \cite{Beringer:1900zz}:
\begin{align}
  m_{\nu_L,3} &= \sqrt{\Delta m_{\rm atom}^2} = 0.048\ {\rm eV},
  \\
  m_{\nu_L,2} &= \sqrt{\Delta m_{\rm solar}^2} = 0.0087\ {\rm eV},
  \\
  m_{\nu_L,1} &\simeq 0.
\end{align}
In addition, $U_{\rm MNS}$ is the so-called Maki-Nakagawa-Sakata (MNS)
matrix, which we parametrize
\begin{align}
  U_{\rm MNS} = &
  \left(
    \begin{array}{ccc}
      c_{12}c_{13} & s_{12} c_{13} & s_{13} e^{-i\delta} \\
      -s_{12} c_{23} -c_{12} s_{23} s_{13} e^{i\delta}
      & c_{12}c_{23} -s_{12}s_{23} s_{13} e^{i\delta} 
      & s_{23} c_{13}\\
      s_{12} s_{23}-c_{12}c_{23}s_{13}e^{i \delta} 
      &-c_{12}s_{23}-s_{12}c_{23}s_{13}e^{i \delta}
      & c_{23}c_{13}
    \end{array}
  \right) 
  \nonumber \\ &
  \times {\rm diag}
  \left(1, \,e^{i\alpha_{21}/2},\,e^{i\alpha_{31}/2}\right),
\end{align}
where $c_{ij} \equiv \cos \theta_{ij}$, $s_{ij} \equiv \sin
\theta_{ij}$.  In our numerical calculation, we take \cite{Beringer:1900zz}
 \begin{align}
   s_{12} = 0.46,~~~ s_{23} = 0.5,~~~ s_{13} = 0.16,
\end{align}
and 
\begin{align}
   \delta = 0.
\end{align}

In addition, in order to capture the most important features of the
present model, we consider the case that the Majorana mass matrix of
right-handed neutrinos has universal structure as\footnote
{As long as $M_{N_3}$ is large enough ($M_{N_3} \sim O(10^{15})~{\rm
    GeV}$), the following discussion is almost unchanged even
  if we choose hierarchical right-handed neutrino masses, $M_{N_{1,2}}
  \ll M_{N_3}$.  Thus, the present discussion is applicable to the
  case compatible with the leptogenesis \cite{Fukugita:1986hr}, which
  requires $M_{N_1} \sim O(10^{9-10})~{\rm GeV}$. }
\begin{align}
  M_{N,IJ} = M_N \delta_{IJ},
\end{align}
and that the neutrino Yukawa matrix is given in the following form:
\begin{align}
  y_{\nu, I j} =
  \frac{\sqrt{2M_N m_{\nu_L, I}} [U_{\rm MNS}]_{Ij}}{v \sin\beta}.
\end{align}
Notice that, with the above assumptions, the parameters $\alpha_{21}$
and $\alpha_{31}$ are irrelevant for the following analysis.

The soft SUSY breaking terms relevant for our following
discussion are given by
\begin{align}
  {\cal L}^{\rm (soft)} = {\cal L}_{\rm MSSM}^{\rm (soft)}
  + m_{\tilde{N}^c,IJ}^2 \tilde{N}^{c*}_I \tilde{N}^c_J
  + (A_{\nu,Ij} \tilde{N}^c_I \tilde{l}_j H_u + {\rm h.c.}),
\end{align}
where ${\cal L}_{\rm MSSM}^{\rm (soft)}$ contains the SUSY breaking
terms of the MSSM.  As we have mentioned, we consider the case where
the SUSY FCNC problems are solved by the assumption of the universal
scalar mass.  In addition, we assume that the tri-linear scalar
coupling constant is proportional to the corresponding Yukawa coupling
constant, like in the case of mSUGRA, at the scale where the boundary
condition is imposed.  Thus, with $\tan\beta$ being fixed, the soft
SUSY breaking parameters related to the scalars are parametrized by
the following parameters:
\begin{align}
  m_0, ~~~ a_0,
\end{align}
where $m_0$ is the universal scalar mass and $a_0$ is the coefficient
of the tri-linear scalar coupling (normalized by $m_0$ as well as by the
corresponding Yukawa coupling constant).  In our analysis, we impose
the boundary condition at the GUT scale $M_{\rm GUT}$, which is taken
to be $2\times 10^{16}\ {\rm GeV}$.

The neutrino Yukawa coupling constants $y_\nu$ change the RG runnings of soft
SUSY breaking parameters, which may affect the low-energy
phenomenology.  In particular, above the mass scale of the
right-handed neutrinos, the soft SUSY breaking mass-squared parameters
of left-handed leptons $m_{\tilde{l},ij}^2$ may be significantly affected.  To see this, it
is instructive to see the RG equations (RGEs) of those parameters; at
the one-loop level, the RGEs above the scale of the right-handed
neutrinos are given by
\begin{align}
  \frac{d m_{\tilde{l},ij}^2}{d \ln Q} = &
  \left[ \frac{d m_{\tilde{l},ij}^2}{d \ln Q} \right]_{\rm MSSM}
  \nonumber \\ &
  + \frac{1}{16\pi^2}
  \left[ 
    (m_{\tilde{l}}^2 y_\nu^\dagger y_\nu + 
    y_\nu^\dagger y_\nu m_{\tilde{l}}^2)
    + 2 (y_\nu^\dagger m_{\tilde{N}^c}^2 y_\nu
    + y_\nu^\dagger y_\nu m_{H_u}^2
    + A_\nu^\dagger A_\nu) \right]_{ij},
  \label{msldot}
\end{align}
where $[d m_{\tilde{l},ij}^2/d\ln Q]_{\rm MSSM}$ denotes the MSSM
contribution to the RGE, $m_{H_u}^2$ is the soft SUSY breaking
mass-squared parameter of up-type Higgs, and $Q$ is the
renormalization scale.  Then, we can easily see that the off-diagonal
elements $m_{\tilde{l},ij}^2$ become non-vanishing at low energy even
if the universality assumption is adopted at the GUT scale.  Indeed,
with the scalar masses being universal at the GUT scale, the
low-energy values of $m_{\tilde{l},ij}^2$
are estimated as
\begin{align}
  m_{\tilde{l},ij}^2 \simeq m_0^2 
  \left[ \delta_{ij}
    - \frac{(y_\nu^\dagger y_\nu)_{ij}}{16\pi^2}
    (6 + 2 a_0^2)
    \ln \frac{M_{\rm GUT}}{M_{N}} \right],
  \label{mhldot}
\end{align}
where we have used the leading-log approximation. As one can see, 
the off-diagonal elements of $m_{\tilde{l},ij}^2$, which become the origin 
of the LFV processes, are generated through the RG effects.  

Another important effect is on the evolution of $m_{H_u}^2$; the RGE
of $m_{H_u}^2$ above the mass scale of the right-handed neutrinos is
\begin{align}
  \frac{d m_{H_u}^2}{d \ln Q} &=
  \left[ \frac{d m_{H_u}^2}{d \ln Q} \right]_{\rm MSSM}
  + \frac{2}{16\pi^2}
  {\rm tr} 
  \left[ y_\nu^\dagger y_\nu m_{H_u}^2
    + y_\nu^\dagger y_\nu m_{\tilde{l}}^2
    + y_\nu y_\nu^\dagger m_{\tilde{N}^c}^2
    + A_\nu^\dagger A_\nu \right].
  \label{mhudot}
\end{align}
The change of the RGE may affect the condition of the electroweak
symmetry breaking, as we discuss below.

In order to discuss the phenomenology at the electroweak scale (and
below), we evaluate the low-energy values of the soft SUSY breaking
parameters by solving the RGEs numerically.  The value of the SUSY
invariant Higgs mass (so-called $\mu$ parameter) is determined by
solving the condition of electroweak symmetry breaking.

Let us first consider the LFVs mainly induced by the off-diagonal
elements of $m_{\tilde{l},ij}^2$.  The values of $m_{\tilde{l},ij}^2$
(with $i\neq j$) are sensitive to model parameters; in particular,
they are approximately linearly dependent on the mass scale of the
right-handed neutrinos with the masses of active neutrinos being
fixed. (See Eq.\ \eqref{seesaw}.)  For the case where the gaugino
masses are equal to $m_0$ at the GUT scale, for example,
$(m_{\tilde{l},12}^2/m_0^2$, $m_{\tilde{l},23}^2/m_0^2$,
$m_{\tilde{l},13}^2/m_0^2$) is about ($-5.0\times 10^{-4}$,
$-1.3\times 10^{-3}$, $-1.5\times 10^{-4}$), ($-3.5\times 10^{-3}$,
$-9.1\times 10^{-3}$, $-1.0\times 10^{-3}$), and ($-1.9\times
10^{-2}$, $-5.0\times 10^{-2}$, $-5.5\times 10^{-3}$) for
$M_N=10^{13}\ {\rm GeV}$, $10^{14}\ {\rm GeV}$, and $10^{15}\ {\rm
  GeV}$, respectively; if all the gaugino masses are much smaller than
$m_0$, we obtain ($-6.8\times 10^{-4}$, $-1.8\times 10^{-3}$,
$-2.0\times 10^{-4}$), ($-4.7\times 10^{-3}$, $-1.2\times 10^{-2}$,
$-1.4\times 10^{-3}$), and ($-2.6\times 10^{-2}$, $-7.9\times
10^{-2}$, $-7.7\times 10^{-3}$) for $M_N=10^{13}\ {\rm GeV}$,
$10^{14}\ {\rm GeV}$, and $10^{15}\ {\rm GeV}$, respectively.  One can
see that the off-diagonal elements become sizable in particular when
$M_N$ is close to $\sim 10^{15}\ {\rm GeV}$, with which the largest
Yukawa coupling constant is $\sim 1$.

To see how large the LFV rates can be, we calculate $Br(\mu\rightarrow
e\gamma)$ in the present setup.  Here, to make our discussion
concrete, we concentrate on two typical models for the choice of the
gaugino masses.
\begin{enumerate}
\item mSUGRA model: If there exists a singlet field in the SUSY
  breaking sector, gaugino masses can dominantly originate from the
  direct interaction between the singlet field and the gaugino.  Then,
  assuming that the interaction respects the GUT symmetry, we
  parametrize
  \begin{align}
    M_A^{\rm (mSUGRA)} (Q=M_{\rm GUT}) = M_{1/2}.
  \end{align}
  The low-energy values of the gaugino masses are determined by
  solving the RGEs with the above boundary condition.
\item Pure gravity mediation model \cite{Ibe:2006de, Ibe:2011aa, ArkaniHamed:2012gw}: 
  If there is no singlet field in the SUSY
  breaking sector, gaugino masses may be dominantly from the effect of
  anomaly-mediated SUSY breaking (AMSB) \cite{Randall:1998uk,
    Giudice:1998xp}; if the pure anomaly-mediation contribution
  dominates, we obtain\footnote
  {If the $\mu$ parameter is as large as $m_0$, there may exist a
    sizable contribution to the gaugino masses via the Higgs-Higgsino
    loop \cite{Giudice:1998xp, Ibe:2006de}.  We neglect such a
    contribution in the present analysis.}
  \begin{align}
    M_{A}^{\rm (AMSB)}
    = -\frac{b_A g_A^{2}}{16\pi^{2}} m_{3/2},
    \label{GauginoMass}
  \end{align}
  where $b_{A}$ denote coefficients of the RGEs of $g_{A}$, i.e.,
  $b_{A}= (-11, -1, 3)$.
\end{enumerate}
In both cases, the scalar masses are assumed to be universal at the
GUT scale.  Requiring successful electroweak symmetry breaking, other
MSSM parameters (i.e., $\mu$- and $B_\mu$-parameters) are determined.
Thus, with $\tan\beta$ being fixed, the low-energy values of the MSSM
parameters are given as functions of $m_0$, $a_0$, ${\rm sign}(\mu)$,
and $M_{1/2}$ or $m_{3/2}$.

\begin{figure}[t]
  \centerline{\epsfxsize=0.8\textwidth\epsfbox{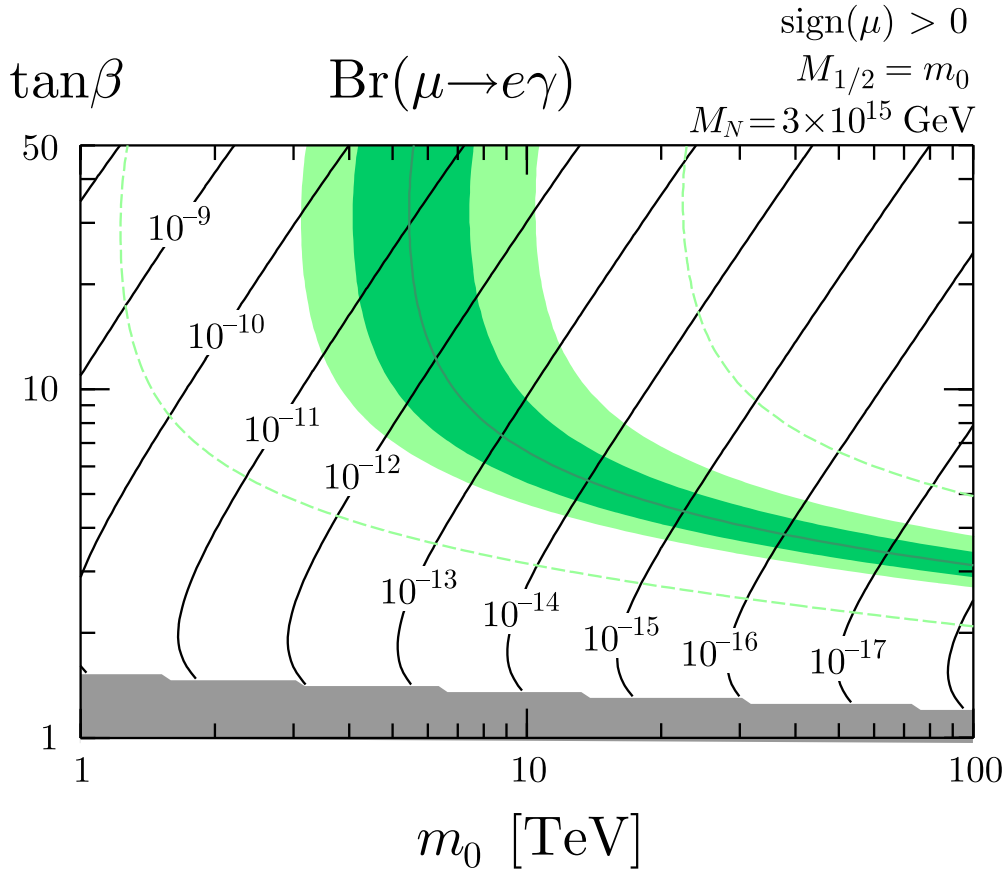}}
  \caption{$Br(\mu\rightarrow e\gamma$) as functions of the universal
    scalar mass $m_0$ and $\tan{\beta}$ for $M_N=3\times 10^{15}\
    {\rm GeV}$, $M_{1/2}=m_0$, $a_0=0$ and ${\rm sign}(\mu)>0$ 
    in the mSUGRA model. 
    Numbers in the figure are the values of ${\rm Br}(\mu \to e \gamma)$.
    Dark (light) green region satisfies $125\ {\rm GeV}<m_h<127\ {\rm GeV} \ 
    (124\ {\rm GeV}<m_h<128\ {\rm GeV})$ and dashed two green lines show 
    $m_h=120\ {\rm GeV}, 130\ {\rm GeV}$. For small $\tan\beta$, gray region 
    is excluded by the non-perturbativity of the top Yukawa coupling constant. }
  \label{fig:mSUGRA1}
\end{figure}

In Fig.\ \ref{fig:mSUGRA1}, we show the contours of constant
$Br(\mu\rightarrow e\gamma)$ on $m_0$ vs.\ $\tan\beta$ plane for the
mSUGRA case.  Here, we take $M_{1/2}=m_0$.  We can see that
$Br(\mu\rightarrow e\gamma)$ can be as large as $O(10^{-13})$ or
larger in the parameter region where the Higgs mass becomes about
$126\ {\rm GeV}$.  With the experimental bound of $Br(\mu\rightarrow
e\gamma)<5.7\times 10^{-13}$ recently reported by the MEG experiment
\cite{Adam:2013mnn}, some of the parameter region (with large
$\tan\beta$) is already excluded even if $m_0$ is as large as $\sim
10\ {\rm TeV}$.  In the future, the MEG upgrade experiment is expected
to improve the sensitivity up to $Br(\mu\rightarrow e\gamma)<6\times
10^{-14}$ \cite{Baldini:2013ke}, which gives better converge of the
parameter space.

As well as the $\mu\rightarrow e\gamma$ process, other LFV reactions
may also occur.  In particular, in the present model, $\mu\rightarrow
3e$ and $\mu$-$e$ conversion processes can be significant.  If the LFV
processes are dominated by the dipole-type operator, which is the case
when $\tan\beta$ is relatively large, $Br(\mu\rightarrow 3e)$ and the
rate of the $\mu$-$e$ conversion $R_{\mu e}$ are both approximately
proportional to $Br(\mu\rightarrow e\gamma)$.  For the $\mu\rightarrow
3e$ process, we obtain $Br(\mu\rightarrow 3e)\simeq 6.6\times
10^{-3}\times Br(\mu\rightarrow e\gamma)$ \cite{Ellis:2002fe}.  The
ratio $R_{\mu e}/Br(\mu\rightarrow e\gamma)$ depends on the nucleus
$N$ used for the conversion process; the ratio $R_{\mu
  e}/Br(\mu\rightarrow e\gamma)$ is approximately given by $2.5\times
10^{-3}$ for $N$ being ${}_{13}^{27}$Al \cite{Kitano:2002mt}.

With the proportionality factor given above, the current bounds on te
$\mu\rightarrow 3e$ and the $\mu$-$e$ conversion processes give less
severe constraints compared to the $\mu\rightarrow e\gamma$ process.
In the future, however, several experiments may improve the bound on
these processes.  For example, the Mu3e experiment may cover
$Br(\mu\rightarrow 3e)\sim 10^{-15}-10^{-16}$ \cite{Blondel:2013ia}.
For the $\mu$-$e$ conversion process, Mu2e \cite{Abrams:2012er} and
COMET \cite{Kuno:2012pt} experiments may reach $R_{\mu e}\sim
10^{-17}$ with $N={}_{13}^{27}$Al, while PRISM/PRIME project
\cite{Kuno:2012pt} may have a sensitivity up to $R_{\mu e}\sim
10^{-19}$.  Thus, these experiments may provide more stringent
constraint on the model of our interest.  In other words, even if all
the superparticles are at the scale of $O(10)\ {\rm TeV}$, the LFV
processes may have large enough rates to be detected at future
experiments.

\begin{figure}[t]
  \centerline{\epsfxsize=0.8\textwidth\epsfbox{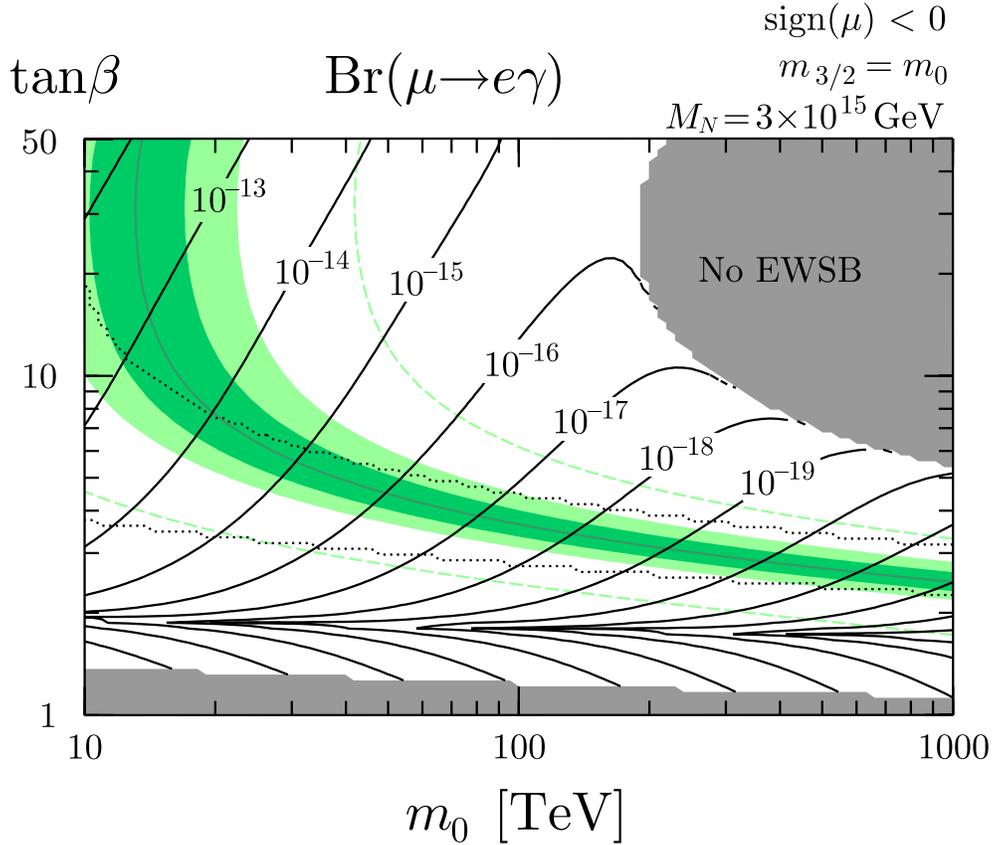}}
  \caption{$Br(\mu\rightarrow e\gamma$) as functions of $m_0$ and
    $\tan{\beta}$ for $M_N=3\times 10^{15}\ {\rm GeV}$ and
    $m_0= m_{3/2}$ in the pure gravity mediation model.  
    Numbers in the figure are the values of ${\rm Br}(\mu \to e \gamma)$.
    Dark (light) green region satisfies $125\ {\rm GeV}<m_h<127\ {\rm GeV} \ 
    (124\ {\rm GeV}<m_h<128\ {\rm GeV})$ and dashed two green lines show 
    $m_h=120\ {\rm GeV}, 130\ {\rm GeV}$. For small $\tan\beta$, gray region 
    is excluded by the non-perturbativity of the top Yukawa coupling constant. 
    For large $\tan\beta$, there is no correct EWSB minimum in the gray region. 
    The upper (lower) dotted lines show the upperbounds
    on $\tan\beta$ by correct EWSB conditions for 
    $M_N = 10^{15}\ {\rm GeV}\ (10^{10}\ {\rm GeV})$.}
  \label{fig:AMSB}
\end{figure}

Next, we consider the situation with smaller gaugino masses; for this
purpose, we adopt the pure gravity mediation model where the gaugino
masses is given by the AMSB and $m_0= m_{3/2}$.  The branching ratio
for the $\mu\rightarrow e\gamma$ process is shown in Fig.\
\ref{fig:AMSB}.  Even in this case, we can see that $Br(\mu\rightarrow
e\gamma)$ can be as large as $O(10^{-13})$ or larger even if we
require that $m_h\simeq 126\ {\rm GeV}$.

In the case of AMSB-type gaugino masses, it should be noted that the
negative searches for the gluino signals at the LHC impose significant
constraint on $m_{3/2}$; $M_3\gtrsim 1.1-1.2\ {\rm TeV}$
\cite{GluinoBound} requires $m_{3/2}\gtrsim 40\ {\rm TeV}$.  If
$m_0=m_{3/2}$, $\tan\beta$ is required to be smaller than $\sim 5$ to
realize $m_h\simeq 126\ {\rm GeV}$; if so, as we can see in Fig.\
\ref{fig:AMSB}, LFV rates are so small that experimental confirmation
of the LFV processes becomes really challenging even in future
experiments. For non-minimal Kahler potential, we do not have to
adopt the relation $m_0=m_{3/2}$; if so we may have a chance to
observe the LFV processes at future experiments even in the case with
AMSB-type gaugino masses.

Next, we discuss the electroweak symmetry breaking in the present
model, because the existence of right-handed neutrinos may have
important effect on it.  In Fig.\ \ref{fig:AMSB}, we can see that the
successful electroweak symmetry breaking can be realized in the region
with large $\tan\beta$; such a region does not exist in the case
without right-handed neutrinos \cite{Evans:2013lpa}.

In fact, it is a generic feature that, with too large universal scalar
mass compared to the gaugino masses, electroweak symmetry breaking
does not occur unless $\tan\beta$ is ${\cal O}(1)$.  This is due to the fact
that, with large $m_0$ and small $M_{1/2}$, it becomes difficult to
realize negative $m_{H_u}^2$, which is essential for the electroweak
symmetry breaking.  Because $m_{H_u}^2>0$ at high scale and also
because the RG running of $m_{H_u}^2$ terminates at the scale of
scalar fermions (which is of the order of $m_0$), $m_0$ should be
small enough to make $m_{H_u}^2$ negative by the RG effect.  In the
present case, $m_{H_u}^2$ is driven to negative by the Yukawa
interactions.  If right-handed neutrinos do not exist, $m_{H_u}^2<0$
is realized by the top Yukawa interaction whose effect is more
enhanced for smaller $\tan\beta$ because the top Yukawa coupling
constant is proportional to $\sim 1/\sin\beta$ (above the mass scale
of superparticles).  As a result, for large $m_0$, smaller value of
$\tan\beta$ is required to have successful electroweak symmetry
breaking if $M_{1/2}$ is relatively small.  (If $M_{1/2}$ is
comparable to $m_0$, the RG effect enhances the stop masses because of
the large gluino mass.  In such a case, the enhanced stop masses make
it easier to realize $m_{H_u}^2<0$.)  If there exist right-handed
neutrinos, the neutrino Yukawa interactions also reduce the low-energy
value of $m_{H_u}^2$, as indicated by Eq.\ \eqref{mhudot}; with the
high-scale (like the GUT-scale) value of $m_{H_u}^2$ being fixed, the
low-scale value of $m_{H_u}^2$ becomes smaller compared to the case
without right-handed neutrinos.  Thus, in models with large scalar
masses, the existence of right-handed neutrinos significantly changes
the condition of the electroweak symmetry breaking.

\begin{figure}[t]
  \centerline{\epsfxsize=0.8\textwidth\epsfbox{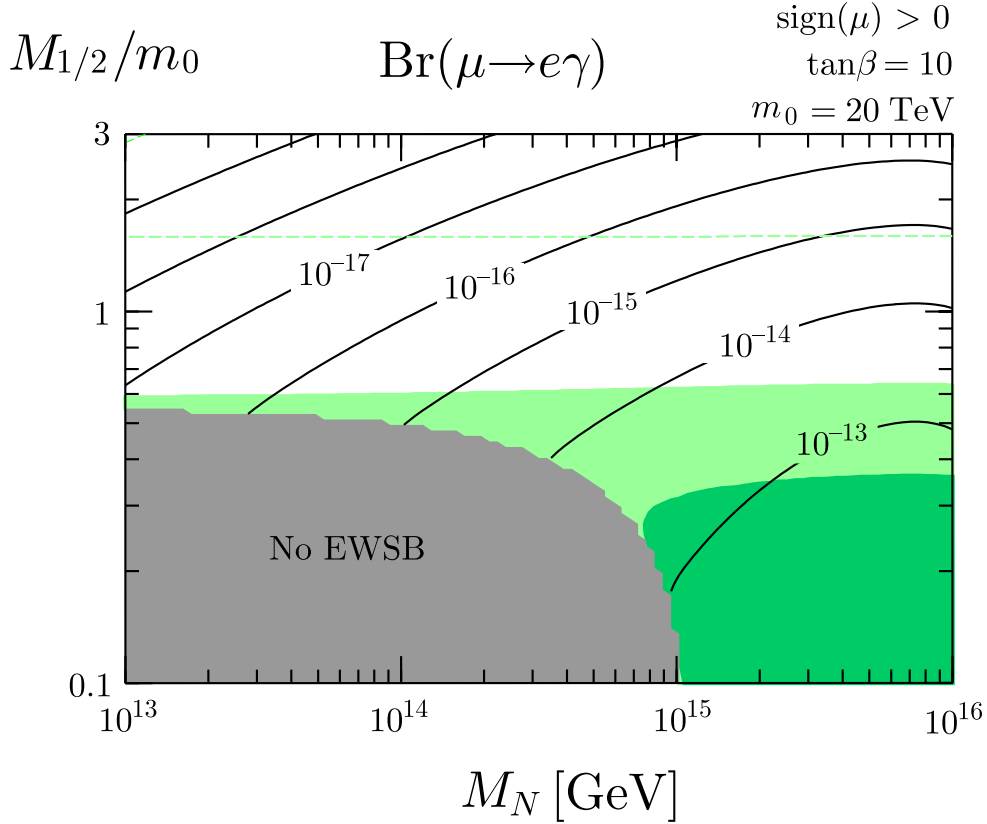}}
  \caption{$Br(\mu\rightarrow e\gamma$) as functions of $M_N$ and
    $M_{1/2}/m_0$ for $m_0=20{\rm TeV}$, $\tan\beta=10$,$a_0=0$ and 
    ${\rm sun}(\mu)>0$ in the mSUGRA model.
    Numbers in the figure are the values of ${\rm Br}(\mu \to e \gamma)$.
    Dark (light) green region satisfies $125\ {\rm GeV}<m_h<127\ {\rm GeV} \ 
    (124\ {\rm GeV}<m_h<128\ {\rm GeV})$ and dashed green line shows 
    $m_h=130\ {\rm GeV}$. For small $M_N$,  there is 
    no correct EWSB minimum in the gray region. }
     \label{fig:mSUGRA2}
\end{figure}

To see the effect of right-handed neutrinos, in Fig.\
\ref{fig:mSUGRA2}, we show the parameter region where electroweak
symmetry breaking successfully occurs in the case of mSUGRA-type
boundary condition on $M_N$ vs.\ $M_{1/2}/m_0$ plane.  (Here, we take
$\tan\beta=10$ and $m_0=20\ {\rm TeV}$.)  We can see that, with larger
value of $M_N$ (corresponding to larger neutrino Yukawa coupling
constant), successful electroweak symmetry breaking becomes possible
in the parameter region with the gaugino masses much smaller than
$m_0$.  This fact indicates that lighter gluino mass becomes allowed
in models with right-handed neutrinos, which makes LHC searches for
superparticles easier.  In the same figure, we also show the contours
of constant $Br(\mu\rightarrow e\gamma)$.  We can see that, in the
region of successful electroweak symmetry breaking newly allowed by
the effect of right-handed neutrinos, the LFV rates can be sizable and
may be within the reach of future experiments.

In summary, we have discussed the LFV rates in SUSY model in which
superparticles (in particular, sfermions) are as heavy as $O(10-100)$
TeV.  The observed Higgs boson mass of 126 GeV suggests the relatively
high scale SUSY breaking with such a mass spectrum. In this letter, we
show that lepton flavor violating processes such as $\mu \rightarrow
e\gamma$ can be in a region accessible to future experiments if the
gaugino masses are of order of gravitino mass and the right-handed
neutrino mass is $O(10^{15})$ GeV suggested by the GUT-like seesaw
mechanism. On the other hand, they are more suppressed in the pure
gravity mediation model and it may be very challenging to observe the
lepton flavor violation in near future experiments.  However, the
gluino mass can be as small as a few TeV in this model which can be
testable at future LHC experiments \cite{Ibe:2012hu, Bhattacherjee:2012ed}.
\footnote{Even without right-handed neutrinos, the gluino mass can be 
  reduced down to a few TeV if there is a saxion flat direction \cite{Nakayama:2013uta}.
  }

\vspace{1em}
\noindent {\it Acknowledgements}: 
T.T.Y. thanks J.Evans, M.Ibe and K.Olive for a useful discussion on the
pure gravity mediation model. This work is supported in part by
Grant-in-Aid for Scientific research from the Ministry of Education,
Science, Sports, and Culture (MEXT), Japan, No.\ 22244021, No.\
22540263, and No.\ 23104008, 
and also by the World Premier International Research
Center Initiative (WPI Initiative), MEXT, Japan.

\end{document}